# The cosmological constant and the gravitational light bending

Tolu Biressa · J.A. de Freitas Pacheco

**Abstract** The solution of the null non-radial geodesic in a Schwarzschild-de Sitter background is revisited. The gravitational bending of a light ray is affected by the cosmological constant, in agreement with the findings of some previous investigations. The present study confirms that the leading correction term depends directly not only on $\Lambda$ but also on the impact parameter and on the angular distance to the source. Using the resulting lens equation, the projected mass of the lens was estimated for several systems displaying Einstein rings. Corrections on the masses due to $\Lambda$ are, on the average, of the order of 2%, indicating that they are not completely negligible for lens systems at cosmological distances.

## 1 Introduction

In the past years a plethora of papers have been addressed to the possible contribution of the cosmological constant $\Lambda$ to the gravitational bending of a light ray passing nearby a massive object. Early studies reported by [1-3] suggested that the bending is not affected by $\Lambda$. The main argument used in these investigations is that in the second order differential equation describing the photon trajectory in polar coordinates, the term containing $\Lambda$ drops out and hence the solution is the same as in a pure Schwarzschild background. As pointed out by Rindler & Ishak [4], this is not exactly the case, since in the presence of a cosmological constant space is not asymptotically flat and this aspect should be taken into account when computing the bending at the observer's position. These authors estimated a first order correction to the ``classical" Schwarzschild bending, which depends on $\Lambda$ but not on the coordinates either of the source or the observer. Subsequent studies [5-6] performed in a expanding background, led to the conclusion that the "classical" bending is practically not affected by the cosmological constant. A further analysis by [7] reached a similar conclusion and the authors interpreted the correction term obtained by [4] as an "aberration" effect caused by the adopted gauge.

Tolu Biressa
University of Addis Ababa
Department of Physics - P.O. Box 1176 - Addis Ababa -Ethiopia
J.A. de Freitas Pacheco
Université de Nice-Sophia Antipolis-Observatoire de la Côte d'Azur
BP 4229, 06304, Nice Cedex 4, France
E-mail: biressa@oca.eu, pacheco@oca.eu

Ishak et al. [8] criticized these studies and reported new calculations in which the bending was now estimated by considering a model where a Schwarzschild-de Sitter vacuole is embedded in a Friedman-Robertson-Walker background. According to them, the correction introduced by the cosmological constant is larger than the second-order term of the "classical" solution.

The first-order differential equation describing the orbit in the plane $\theta = \pi/2$, can be solved in terms of elliptical functions ([9]) and, more recently, Sereno [10] obtained approximate solutions by using a convenient series expansion of those solutions. He obtained high order corrections to the "classical" bending angle, taking into account the fact that the space-time is not asymptotically flat. The leading correction term obtained by him is of the form $2Mb\Lambda/3$, depending directly on the impact parameter $b$, on the lens mass M and on the cosmological constant. This term was interpreted by [11] as a "coupling" between the lens and $\Lambda$, labelled "local" by the author, since it does not depend on the positions either of the source or the observer.

In the present work a new investigation on this problem is reported. The solution of the orbit equation in the first order of the small parameter $3M/r_0$, where $r_0$ is the closest distance approach is revisited, taking into account adequate boundary conditions. The bending was calculated, as it should, at the observer's position and not at an arbitrary position of the orbit as some past authors did. The leading correction term of our solution depends directly not only on the $\Lambda$ term but also on the distance between the observer and the source. The resulting lens equation was used to estimate masses of some stellar systems and our computations indicate that the contribution due to the cosmological constant may introduce corrections of about 2% in masses, for lens systems situated at z ~ 1. The paper is organized as follows: in Section 2 the equations are presented, in Section 3 the lens equation is obtained and applied to some stellar systems associated to almost circular gravitational arcs and, finally, in Section 4 the main results are summarized.

**2 Solution of the orbit equation**

The background space-time considered here is defined by the well-known Schwarzschild-de Sitter (SdS) metric

$$ds^2 = -f(r)dt^2 + f^{-1}(r)dr^2 + r^2 d\Omega^2 \quad (1)$$

where

$$f(r) = 1 - \frac{2M}{r} - \frac{\Lambda r^2}{3} \quad (2)$$

and geometric units were used. The mass $M$ represents the lens situated at the origin of the coordinate system. If the motion is assumed to occur in the plane $\theta = \pi/2$, the non-radial null geodesic equation associated with eqs.1 and 2 is

$$\left(\frac{1}{r^2}\frac{dr}{d\phi}\right)^2 = \frac{1}{b_\Lambda^2} - \frac{1}{r^2} + \frac{2M}{r^3} = \frac{1}{b^2} - \frac{f(r)}{r^2} \quad (3)$$

where we have introduced the effective impact parameter $b_\Lambda$ by the relation

$$\frac{1}{b_\Lambda^2} = \frac{1}{b^2} + \frac{\Lambda}{3} \quad (4)$$

Here $b$ is the usual impact parameter defined by the ratio between the specific angular momentum and energy, both being constants of motion. The closest approach distance $r_0$ is derived by imposing the condition $dr/d\phi = 0$ in eq. 3. The resulting equation is an algebraic equation of the third degree in $r_0$, whose root of interest (the minor real root) is given by

$$\frac{r_0}{b_\Lambda} = \frac{2}{\sqrt{3}} \cos\left[\frac{1}{3}\cos^{-1}\left(-\frac{\sqrt{27}M}{b_\Lambda}\right)\right] \tag{5}$$

The equation above is exact bur since in general the condition $2M/b_\Lambda \ll 1$ is satisfied, a series expansion of this equation gives

$$\frac{r_0}{b_\Lambda} \approx 1 - \frac{M}{b_\Lambda} - \frac{3}{2}\left(\frac{M}{b_\Lambda}\right)^2 + ... \tag{6}$$

In terms of the dimensionless variable $u = r_0/r$, eq. 3 can be recast as

$$\left(\frac{du}{d\phi}\right)^2 = \left(\frac{r_0}{b_\Lambda}\right)^2 - u^2 + \frac{2M}{r_0}u^3 \tag{7}$$

Performing the derivative of both sides of this equation with respect to the coordinate $\phi$, one obtains the second-order equation

$$\frac{d^2u}{d\phi^2} + u = \alpha u^2 \tag{8}$$

where we have defined the parameter $\alpha = 3M/r_0 \ll 1$. Using the well known perturbation procedure to obtain an approximate solution of eq. 8 in terms of the small parameter $\alpha$, on can write

$$u(\phi) = u_0(\phi) + \alpha u_1(\phi) + \alpha^2 u_2(\phi) + .... \tag{9}$$

Substituting this series in eq. 8, ordering the terms of same power in the perturbation parameter $\alpha$ and keeping only up to the first order terms, one obtains the following set of equations

$$\frac{d^2u_0}{d\phi^2} + u_0 = 0 \tag{10}$$

and

$$\frac{d^2u_1}{d\phi^2} + u_1 = u_0^2 \tag{11}$$

The general solution of eq. 10 is trivial and is given by

$$u_0 = a_0 \cos\phi + b_0 \sin\phi \tag{12}$$

Introducing this equation into eq. 11 and solving, one obtains

$$u_1 = \frac{1}{2}\left(a_0^2 + b_0^2\right) - A_1 \sin\phi + A_2 \cos\phi + \frac{1}{6}\left(b_0^2 - a_0^2\right)\cos 2\phi - \frac{1}{3}a_0 b_0 \sin 2\phi \tag{13}$$

In order to determine the integration constants, we impose the following conditions: since the closest approach occurs at $\phi = \pi/2$, we must have $u(\pi/2) = 1$, which is equivalent to impose $u_0(\pi/2) = 1$ and $u_1(\pi/2) = 0$. From these last two conditions one obtains respectively $b_0 = 1$ and $A_1 = \left(1 + 2a_0^2\right)/3$. Moreover, at the closest approach position, the orbit equation must satisfy $du/d\phi = 0$ and, splitting again this condition into $du_0/d\phi = 0$ and $du_1 d\phi = 0$ it results $a_0 = 0$ and $A_2 = 0$. Thus, at first order, the orbit equation becomes

$$u = \sin\phi + \frac{\alpha}{2}\left(1 + \frac{1}{3}\cos 2\phi - \frac{2}{3}\sin\phi\right) \tag{14}$$

It can be trivially verified that eq. 14 satisfies all the required conditions. In the case of a Schwarzschild space-time, the total bending angle can be estimated from the angle between the two asymptotes at the limit $r \to \infty$, since space is asymptotically flat. This is not the case for a SdS space-time, in which an observer dependent horizon is present. In order to compute the bending angle we follow the approach adopted in [4], using the invariant formula for the tangent of the angle $\psi$ between two coordinate directions, i.e., the radial and the tangent to the orbit at a given point P, which is given by

$$\tan\psi = \frac{\sqrt{g_{\phi\phi}}}{\sqrt{g_{rr}}}\left|\frac{d\phi}{dr}\right| = r\sqrt{f(r)}\left|\frac{d\phi}{dr}\right| \tag{15}$$

Using eq. 3, the equation above can be recast as

$$\tan\psi = \frac{\sqrt{f}}{\sqrt{\frac{r^2}{b^2} - 1}} \tag{16}$$

Since the impact parameter can be expressed in terms of the closest approach distance $r_0$ by the relation $b^2 = r_0^2/f(r_0)$, the equation above can be written as

$$\tan\psi = \frac{1}{\sqrt{\frac{f_0 r^2}{f r_0^2} - 1}} \tag{17}$$

where $f_0 \equiv f(r_0)$. Developing the above equation one obtains

$$\tan\psi = \sqrt{\frac{\left[\frac{r_0^2}{r^2}\left(1 - \frac{r_g}{r}\right) - \frac{\Lambda r_0^2}{3}\right]}{\left[1 - \frac{r_0^2}{r^2} - \frac{r_g}{r_0}\left(1 - \frac{r_0^3}{r^3}\right)\right]}} \tag{18}$$

Since the angle $\psi$ is quite small for distances $r \gg 2M$ and $r \gg r_0$, the eq. 18 can be expanded in series and one obtains

$$\psi \approx \frac{r_0}{r} + \frac{M}{r} - \frac{Mr_0}{r^2} - \frac{\Lambda r_0 r}{6} - \frac{\Lambda r_0^3}{12r} - \frac{\Lambda Mr}{6} \tag{19}$$

This series expansion agrees with that obtained in reference [12] excepting that their last term, corresponding to our fifth term, is twice the value here obtained.

The bending angle $\delta$ at a given orbital angular position $\phi$ is simply $\delta = \psi - \phi$ and the total bending is obtained by adding the contributions at source and the observer positions, i.e.,

$$\delta = (\psi_S + \psi_{obs}) - (\phi_S + \phi_{obs}) \tag{20}$$

Notice that the choice of the points on the orbit at which the bending is estimated has been somewhat arbitrary in past investigations. In [4], for instance, the observer was assumed to be at $\phi = 0$ and the corresponding $r$ coordinate was derived from the orbit equation. For the coordinate system adopted here (polar axis coincident with the direction of the closest approach), the position of the observer (or the source) cannot be taken as zero (or $\pi$), as also emphasized by [12]. Now taking the orbit equation (eq. 14}) at the observer's position ($\phi_{obs}$, $r_{obs}$) and at the source position ($\phi_S$, $r_S$) and performing a series expansion around $\phi_S \sim 0$ and $\phi_{obs} \sim \pi$, one obtains

$$(\phi_S + \phi_{obs}) \approx -\frac{4M}{r_0} + r_0 \Delta_1 + M \Delta_1 \tag{21}$$

where have defined for short

$$\Delta_1 = \left(\frac{1}{r_S} + \frac{1}{r_{obs}}\right) \tag{22}$$

$$\Delta_2 = \left(\frac{1}{r_S^2} + \frac{1}{r_{obs}^2}\right) \tag{23}$$

with $r_S$ and $r_{obs}$ being respectively the radial coordinates of the source and the observer.

From eqs. 19, 20 and 21 one obtains

$$\delta \approx \frac{4M}{r_0} - \frac{\Lambda r_0}{6}(r_S + r_{obs}) - \frac{\Lambda M}{6}(r_S + r_{obs}) - \frac{\Lambda r_0^3}{12}\Delta_1 - Mr_0\Delta_2 + ... \tag{24}$$

Using now eqs. 4 and 6 to express the closest approach distance in terms of the impact parameter $b$ and of the cosmological constant, one obtains finally for the bending

$$\delta \approx \frac{4M}{b} - Mb\Delta_2 + \frac{2Mb\Lambda}{3} - \frac{b\Lambda}{6}(r_S + r_{obs}) - \frac{b^3\Lambda}{12}\Delta_1 + \frac{Mb^3\Lambda}{6}\Delta_2 + ... \tag{25}$$

This result should be compared with the recent study by [12]. Notwithstanding the differences in the derived solution of the orbit equation, the resulting bending angles are quite similar. The first, the third and the fourth terms are the same in both solutions but there are some differences in the second term (twice the value here obtained) and in the fifth, which in their solution has the opposite sign and is two times bigger. Notice that now the main correction to the ``classical'' bending is the fourth term of eq. 25, not present in the results by [4], [10], [11] but present in the work by [12].

**3 The lens equation**

If $\beta$ and $\theta$ are respectively the angular positions of the source and the image relative to the lens direction then, under the assumption that $\beta$ and $\theta$ are small angles, the lens equation is simply

$$\beta = \theta - \delta(\theta)\frac{D_{LS}}{D_S} \quad (26)$$

where $D_{LS}$ and $D_S$ are respectively the angular distances between the lens and the source and between the observer and the source. In this equation, the cosmological constant affects not only $\delta(\theta)$ but also the angular distances. Moreover, as pointed out in [11], the cosmological constant affects also the redshift of the images, but this correction will be not considered here.

Angular distances are measurable quantities, which can be expressed as a function of the redshift for a given cosmological model. This is not the case of eq. 25, in which different terms are expressed as functions of (``static'') radial coordinates. Relations between these coordinates and angular distances were derived in [13] and are given by

$$r_{obs} = D_L \quad (27)$$

and

$$\frac{(r_S + r_{obs})}{\sqrt{1 - \frac{\Lambda r_S^2}{3}}} = D_S \quad (28)$$

where $D_L$ is the angular distance to the lens.

Using these relations, the lens equation can be recast as

$$\beta = \theta - \frac{4M}{b}\frac{D_{LS}}{D_S} - \alpha_\Lambda(\theta)\frac{D_{LS}}{D_S} \quad (29)$$

where the bending angle was split into the main or ``classical'' contribution (first term of eq. 25) and corrections introduced by the cosmological constant, including curvature effects in the measurement process (all the remaining terms of eq. 25). Restricting our analysis only to the leading correction terms (first order in $\Lambda$ and in $M$) one obtains

$$\alpha_\Lambda(\theta) \approx -\frac{\Lambda}{6}D_S D_L \theta + \frac{2M\Lambda}{3}D_L \theta \quad (30)$$

## 3.1 The effect of $\Lambda$ on mass estimates

If the lens is an extended object, the total light deflection is given by the integral of the contribution of each mass element. In the majority of the cases, the deviation of the light ray from the original path during the crossing of the lens is quite small. In this case, we have a ``thin'' lens and its mass density distribution can be projected into a surface density on a plane perpendicular to the line of sight throughout the lens. The angular separation of the image can be estimated from eq. 29 but one should replace the point source M by the projected mass inside the impact parameter *b*.

It should be emphasized that it is not our aim to develop here a detailed model of the mass distribution of an extended lens. Our goal is only to quantify the effects of the corrections introduced by the cosmological constant by comparing the resulting lens masses with and without such corrections.

With such a warning, two possible situations are examined. The first, in which besides the knowledge of Einstein's radius as well as the redshift of the source and the lens, the central velocity dispersion is also known. In this case, we assume that the mass density profile of the lens is modeled by a non-singular profile given by the relation

$$\rho(r) = \frac{A}{\left(r_c^2 + r^2\right)} \qquad (31)$$

where *A* is a constant and $r_c$ is the core radius. Using the equation above, the projected mass inside the impact parameter *b* is calculated trivially and is given by

$$M(<b) = 2\pi^2 A \left[ \sqrt{\left(r_c^2 + b^2\right)} - r_c \right] \qquad (32)$$

In order to relate the constant A to the stellar velocity dispersion $\bar{\sigma}$ in the line-of-sight, we suppose that the lens is in dynamical equilibrium described by the Jeans equation in spherical symmetry and with an isotropic velocity dispersion distribution. Under these conditions

$$\frac{d(\rho\sigma^2)}{dr} = \rho\nabla\Psi = -\rho\frac{GM(r)}{r^2} \qquad (33)$$

Using eq. 31 and imposing the condition $\lim(\rho\sigma^2) = 0$ for $r \to \infty$, the Jeans equation can be integrated and one obtains

$$\sigma^2(r) = \frac{4\pi GA}{r_c^2}\left(r^2 + r_c^2\right)\left[\frac{\pi^2}{8} - \frac{r_c}{r}\arctan\left(\frac{r}{r_c}\right) - \frac{1}{2}\arctan^2\left(\frac{r}{r_c}\right)\right] \qquad (34)$$

Notice that the velocity dispersion is not a constant as in the singular isothermal model. Observers measure the velocity dispersion in the line-of-sight weighted by the stellar light or the stellar mass if the mass-to-luminosity ratio is assumed to be constant. Thus, the observed velocity dispersion is given by the average $\bar{\sigma}^2 = \int \rho\sigma^2 dr / \int \rho dr$. We have performed numerically these integrals using eqs. 31 and 34 in the interval $0 \leq r \leq 100 r_c$ and, from these computations

$$\bar{\sigma}^2 \approx 1.222\pi GA \qquad (35)$$

Substituting this result in eq. 32 one obtains

$$M(<b) = 1.637 \frac{\pi \bar{\sigma}^2}{G}\left[\sqrt{(r_c^2 + b^2)} - r_c\right] \qquad (36)$$

Considering the particular case in which the source is practically aligned with the lens ($\beta \simeq 0$), situation in which the image constitutes a ring whose angular radius $\theta_E$ is obtained by replacing eq. 36 into eq. 29, i.e.,

$$\theta_E^2 = \frac{B}{u(1+F_c)}\left[\sqrt{1+u^2\theta_E^2} - 1\right] \qquad (37)$$

where we have defined the dimensionless variable $u = D_L / r_c$ and the parameter

$$B = 6.547\pi\left(\frac{\bar{\sigma}}{c}\right)^2 \frac{D_{LS}}{D_S} \qquad (38)$$

The effect of the cosmological constant on the bending is represented by the term $F_c$, given to the first order correction by

$$F_c \approx \frac{\Lambda}{6} D_L D_{LS} - \frac{2}{3}\frac{GM(<b)}{c^2}\frac{\Lambda D_L D_{LS}}{D_S} \qquad (39)$$

From inspection of the Cfa-Arizona Space Telescope LEns Survey (CASTLES) of gravitational lenses [14], three ring systems were selected whose properties are given in table 1.

Table 1 Properties of ring lens systems. The first column identifies the object, the second and the third columns give the redshift of the lens and source respectively, the fourth column gives the measured angular radius of the ring and the last column gives the central velocity dispersion of the lens.

| Object | $z_L$ | $z_S$ | $\theta_E$ (arcsec) | $\sigma$ (km/s) |
|---|---|---|---|---|
| Q0047-2808 | 0.48 | 3.60 | 1.4 | 229 |
| CFRS03.1077 | 0.94 | 2.94 | 1.1 | 256 |
| MG1549+3047 | 0.11 | 1.17 | 0.9 | 227 |

Numerical solution of eq. 37 permits to obtain for each object the core radius, which inserted into eq. 36 allows an estimation of the lens projected mass inside the impact parameter $b$. For each object listed in table 1, two solutions were obtained: one imposing $F_c = 0$, i.e., without the correction due to the cosmological constant and another in which $F_c$ is given by eq. 39. Results are shown in table 2. Corrections are of the order of 2% and increases slightly as the redshift of the lens increases. In all computations we consider a flat cosmological model defined by the parameters $\Omega_\Lambda = 0.73$, $\Omega_m = 0.27$ and $H_0 = 71$ $km/s/Mpc$.

**Table 2** Core radius, impact parameter (in kpc) and projected lens masses (in units of $10^{11}\ M_\odot$) for ring systems. The first line corresponds to values derived with $F_c = 0$ while the second line corresponds to values including the correction due to $\Lambda$.

| Object | $r_c$ (kpc) | b (kpc) | $M_p(<b)$ |
|---|---|---|---|
| Q0047-2808 | 2.486 | 8.03 | 3.589 |
|  | 2.722 | 8.03 | 3.490 |
| CFRS03.1077 | 3.400 | 8.31 | 4.223 |
|  | 3.644 | 8.31 | 4.110 |
| MG1549+3047 | 1.779 | 1.69 | 0.400 |
|  | 1.810 | 1.69 | 0.395 |

When the velocity dispersion is not available, it is possible to model the mass distribution of the lens by a singular isothermal profile given by

$$\rho(r) = \frac{\sigma^2}{2\pi G r^2} \qquad (40)$$

In this case, the projected mass inside the impact parameter can be expressed as a function only of the angular ring radius $\theta_E$ by the relation

$$M(<b) = \frac{c^2}{4G}\frac{D_S D_L}{D_{LS}}(1+F_c)\theta_E^2 \qquad (41)$$

while the velocity dispersion in the line of sight can be estimated from the equation

$$\sigma = \frac{c}{2\sqrt{\pi}}\sqrt{\frac{D_S}{D_{LS}}(1+F_c)\theta_E} \qquad (42)$$

These equations were applied to five bona fide gravitational rings taken from the recent COSMOS survey [15] and the results are given in table 3. As in the precedent case, corrections due to the cosmological constant on the mass values are of the order of 2%. Notice that the effect introduced by $\Lambda$ in the estimate of the lens mass depends on the adopted model. Masses are slightly lower if a cored density profile is used and slightly higher if an isothermal density profile is adopted.

**4 Conclusions**

A more general approximate solution of the non radial null geodesics in a SdS space-time was obtained, whose integration constants were fixed by adequate conditions, obliging in particular the orbit to pass by the source position. Using the same procedure adopted in [4], the bending angle was estimated at the observer's position, taking into account the fact that the SdS space-time is not asymptotically flat.

Table 3 Lens parameters for the singular isothermal model. The first column identifies the object, the three following columns give respectively the redshift of the lens, that of the source and the angular radius of the ring. The fourth column gives the impact parameter, the fifth column gives the derived velocity dispersion and the last column the projected mass inside the impact parameter in units of $10^{11}\ M_\odot$. As in table 2, the second line corresponds to values including the correction due to the cosmological constant.

| Object | $z_L$ | $z_S$ | $\theta_E (arcsec)$ | b (kpc) | $\sigma$ (km/s) | $M_p(<b)$ |
|---|---|---|---|---|---|---|
| 0038+4133 | 0.89 | 2.70 | 0.72 | 5.61 | 222 | 1.958 |
|  |  |  |  |  | 225 | 2.012 |
| 0049+5128 | 0.33 | 0.74 | 1.07 | 5.05 | 271 | 2.620 |
|  |  |  |  |  | 273 | 2.660 |
| 0124+5121 | 0.84 | 2.47 | 0.24 | 1.84 | 128 | 0.213 |
|  |  |  |  |  | 130 | 0.219 |
| 5941+3628 | 0.90 | 2.76 | 0.76 | 5.94 | 228 | 2.184 |
|  |  |  |  |  | 231 | 2.245 |
| 5947+4752 | 0.28 | 0.61 | 0.51 | 2.15 | 188 | 0.535 |
|  |  |  |  |  | 189 | 0.542 |

The cosmological constant affects the orbit, introducing corrections to the ``classical'' bending in agreement with some recent investigations [4], [10], [8], [11]. The leading correction terms differ among these investigation, a consequence of the choice of the observer's position, not always consistent with the choice of the orientation of the adopted polar frame. The present investigation confirms the result derived in reference [12], in which the leading correction term is of the form $-\Lambda b D_S / 6$ and which gives a non negligible contribution for sources at cosmological distances.

By using simple mass distribution models for the lens, the projected mass inside the impact parameter $b$ was estimated for several systems displaying a ring image, case in which the source is practically aligned with the lens. Corrections introduced by the cosmological constant produce variations on the average of the order of 2% in the mass values and increase slightly for systems located at higher redshifts, whereas they are completely negligible at solar system scales.

**Acknowledgements**  TB thanks the Observatoire de la Côte d'Azur for its hospitality during the preparation of this work. The authors also thank the referees of this paper for their useful comments.